\documentclass{aa}  

\usepackage{graphicx}
\usepackage[varg]{txfonts}
\usepackage[colorlinks=true,allcolors=blue,backref]{hyperref}
\begin{document}

   \title{A test of Ca {\sc ii} H \& K photometry for isolating massive globular clusters below the metallicity floor}
   \subtitle{}
   \author{
   Bas van Heumen\inst{1}\thanks{Based on observations obtained with MegaPrime/MegaCam, a joint project of CFHT and CEA/DAPNIA, at the Canada-France-Hawaii Telescope (CFHT) which is operated by the National Research Council (NRC) of Canada, the Institut National des Science de l'Univers of the Centre National de la Recherche Scientifique (CNRS) of France, and the University of Hawaii. The observations at the Canada-France-Hawaii Telescope were performed with care and respect from the summit of Maunakea which is a significant cultural and historic site.}
   \and
   William E. Harris\inst{2}
   \and
   S\o ren S. Larsen \inst{1}
   \and
   Else Starkenburg\inst{3}
   }
   \institute{Department of Astrophysics/IMAPP, Radboud University, PO Box 9010, 6500 GL, The Netherlands
   \and
   Department of Physics \& Astronomy, McMaster University, 1280 Main Street West, Hamilton, L8S 4M1, Canada
   \and
    Kapteyn Astronomical Institute, University of Groningen, Landleven 12, 9747 AD Groningen, The Netherlands
   }
   \date{Received XXX; accepted XXX}

 
  \abstract
   {The serendipitous discovery of the M31 globular cluster (GC) EXT8 has presented a significant challenge to current theories for GC formation. By finding other GCs similar to EXT8, it should become clear if and/or how EXT8 can fit into our current understanding of GC formation.}
   {We aim to test the potential of integrated-light narrow-band Ca {\sc ii} H \& K photometry as a proxy for the metallicity of GCs to be able to provide effective candidate selection for massive GCs below the GC metallicity floor ([Fe/H] $\leq$ -2.5), such as EXT8.}
   {We investigate the behaviour of two colours involving the CaHK filter employed by the \textit{Pristine} survey, CaHK-u and CaHK-g, as a function of metallicity through CFHT MegaCam imaging of EXT8 and a wide set of M31 GCs covering the metallicity range of -2.9 $\leq$ [Fe/H] $\leq$ +0.4. Additionally, we investigate if the CaHK colours are strongly influenced by horizontal branch morphology through available morphology measurements.}
   {In both of the CaHK colours, EXT8 and two other potential GCs below the metallicity floor can be selected from other metal-poor GCs ([Fe/H] $\leq$ -1.5) with $(CaHK-g)_o$ showing the larger metallicity sensitivity. The RMS of linear fits to the metal-poor GCs show an uncertainty of 0.3 dex on metallicity estimations for both colours. Comparisons with u-g and g-z/F450W-F850L colours reinforce the notion that CaHK photometry can be used for effective candidate selection as they reduce false positive selection rates by at least a factor of 2. We find no strong influence of the horizontal branch morphology on the CaHK colours that would interfere with candidate selection, although the assessment is limited by quantity and quality of available data.}
   {}
   {}

   \maketitle
   

\section{Introduction} \label{Sec:Intro}
Observations of Galactic and extragalactic globular clusters (GCs) have shown a lack of GCs with [Fe/H] $\leq$ -2.5 with respect to the expectations from chemical evolution models or the metallicity distribution of halo stars (e.g. \citealt{Bond1981,Carney1996, Youakim2020}). The halo metallicity distribution of \cite{Youakim2020} suggests that not observing a GC with [Fe/H] $\leq$ -2.5 in the Galaxy is around a 0.5 percent chance. This points to a genuine truncation in the GC metallicity function and has resulted in the notion of a minimum metallicity for GCs, referred to as the GC metallicity floor at [Fe/H]$_{\text{min}}$ $\approx$ -2.5 \citep{Forbes2018,Beasely2019}.

The origin of the metallicity floor remains unclear. A simple explanation is that GCs are unable to form in extremely low metallicity environments. Possible culprits are the gas fragmentation properties or inefficient cooling expected in such environments (e.g. \citealt{Loeb1994,Abel2002}). A parallel can be drawn to the G-dwarf problem \citep{vandenBergh1962} as well, suggesting that the ISM might have been already sufficiently enriched when GCs started forming. It has also been suggested that the metallicity floor is a natural consequence of the galaxy mass-metallicity relation and the hierarchical nature of galaxy assembly \citep{Kruijssen2019}.

In the latter scenario, the metallicity floor arises as galaxies with [Fe/H] $\leq$ -2.5 are not massive enough to form GCs with sufficient mass to avoid total dissolution due to various dynamical effects for up to a Hubble time (estimated at 10$^5$ M$_{\odot}$; \citealt{Campos2018}). This scenario is supported by the recent discoveries of two stellar streams, Phoenix ([Fe/H] = -2.7 $\pm$ 0.06; \citealt{Wan2020}) and C-19 ([Fe/H] = -3.38 $\pm$ 0.26; \citealt{Martin2022}), which could represent remnants of low-mass ($\sim$ 10$^4$ M$_{\odot}$) GCs. However, we note that the total initial stellar mass of C-19 is uncertain. The rather large observed velocity dispersion and stream width leave a range of possibilities for its progenitor properties (see \citealt{Errani2022,Yuan2022,Viswanathan2024,Yuan2025,Carlberg2025}).\\
\\
Another recent discovery related to the metallicity floor is a serendipitous finding by \cite{Larsen2020}. Using a high-S/N and high-resolution spectrum, the M31 GC EXT8 was found to have a metallicity of [Fe/H] = -2.91 $\pm$ 0.04 and a mass of 1.1 $\times$ 10$^6$~M$_{\odot}$. This combination of extremely low metallicity and high mass is in such stark contrast to the expected combinations from current theory that EXT8 should not exist. A colour-magnitude diagram of EXT8 obtained with HST confirmed that it is a genuine old but very metal-poor GC \citep{Larsen2021}. EXT8 and the associated implied existence of massive GCs below the metallicity floor presents a strong challenge to the currently most accepted paradigm of GC formation \citep{Kruijssen2015}. 

Such a serendipitous discovery naturally calls for a more structured search to improve the statistics and thus put tighter constraints on possible scenarios for the existence of massive GCs below the metallicity floor. Finding more GCs like EXT8 will, however, be very challenging given the combination of their observed rarity and extremely low metallicity. While accurate metallicity determinations at [Fe/H] $\leq$ -2 necessitate high-S/N and high-resolution spectra, such observations are expensive in terms of the amount of observing time required at current facilities, and an effective way of pre-selecting the most metal-poor candidates is therefore desirable. Traditional broad-band photometry becomes increasingly insensitive to metallicity in this regime, but narrow-band photometry employing carefully chosen filters may provide an effective alternative to more time consuming spectroscopic observations.\\
\\
Indeed this approach has proven quite successful in the search for the most metal-poor individual stars. In particular, using the Ca {\sc ii} H \& K lines for candidate selection has resulted in some of the most successful surveys, such as the HK and Hamburg-ESO \citep{Beers1992, Christlieb2008} surveys. The Ca {\sc ii} H \& K lines even continue to be utilized in modern surveys, such as \textit{SkyMapper} and \textit{Pristine} \citep{Keller2008, Starkenburg2017}. Specifically, \textit{Pristine} uses a narrowband filter centred on the Ca {\sc ii} H \& K lines, named CaHK, in combination with SDSS or Gaia filters to be able to distinguish metallicities down to [Fe/H] $\sim$ -3 and has reported a purity of 77 percent for their selection of stars with [Fe/H] $\leq$ -2.5 \citep{Viswanathan2024}.

This success stems from the fact that the Ca {\sc ii} H \& K lines are quite prominent and retain their prominence far better than other metal lines in late-type stars as metallicity decreases (see Fig. 1 from \citealt{Starkenburg2017}). This allows the Ca {\sc ii} H \& K lines to serve as a proxy for metallicity that retains good sensitivity even in the very and extremely metal-poor regimes. 

As GCs are nearly mono-metallic and late-type stars constitute a majority of their stellar population, it tentatively suggests that the Ca {\sc ii} H \& K lines could also be used as a metallicity indicator to identify very metal-poor GCs.

This potential of Ca {\sc ii} H \& K lines has already been noted by \cite{Zinn1980}. who upon studying an spectrum of M92 ([Fe/H] = -2.31 dex; \citealt[2010 edition]{Harris1996}) remarked that the only strong metal lines remaining were the Ca {\sc ii} H \& K lines and they might be the best, if not the only, option of distinguishing a GCs more metal-poor than M92 from M92 through integrated-light (IL).\\
\\
With the serendipitous discovery of EXT8 and the challenge it poses to the current ideas about GC formation, the potential of the Ca {\sc ii} H \& K lines as metallicity indicator for metal-poor GCs could be an important tool to make the search for massive GCs below the metallicity floor the most efficient it can be. 

This study aims to test this potential of the Ca {\sc ii} H \& K lines for GCs by investigating whether EXT8 could be identified through Ca {\sc ii} H \& K photometry using the CaHK filter from \textit{Pristine}. To this end, EXT8 alongside a wide sample of GCs in M31 were imaged in \textit{CaHK}, \textit{u}, \textit{g} and \textit{i} by MegaCam from the Canada-France-Hawaii Telescope (CFHT).

\section{Data} \label{Sec:Data}
\begin{figure}
\centering
\includegraphics[width=\hsize]{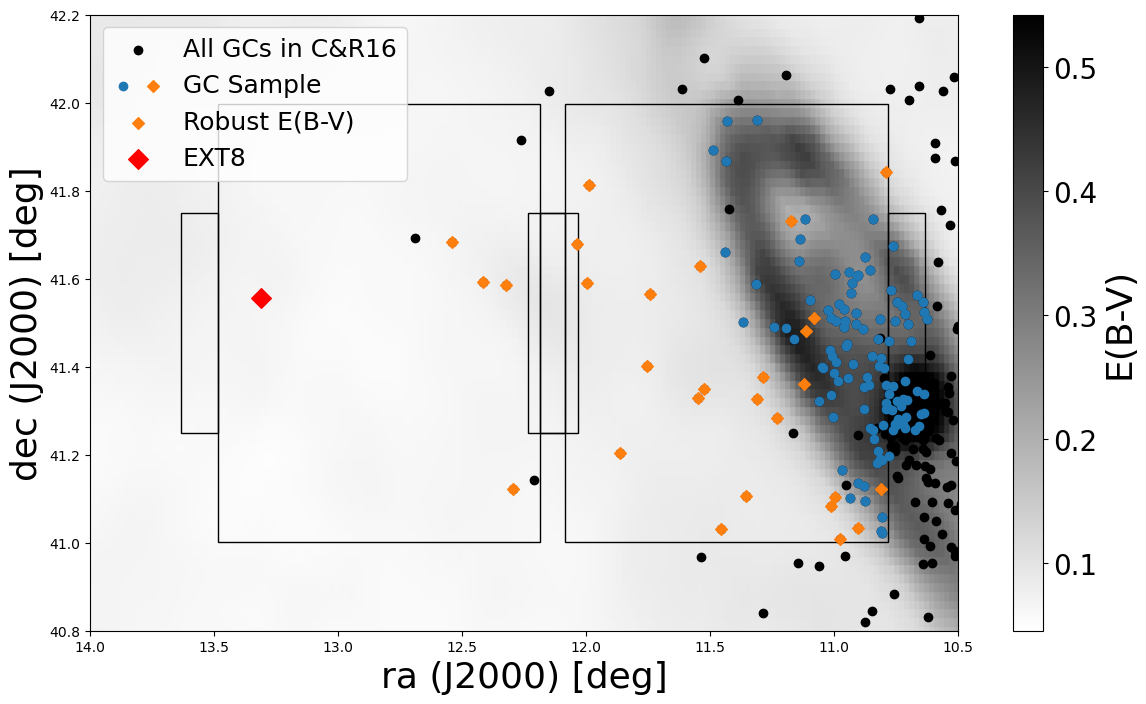}
    \caption{The observational footprint of this study. The GC sample has been marked with either orange diamonds or blue circle, signifying GCs with and without robust reddening estimates. EXT8 has been marked in red and the black circles are confirmed M31 GCs from the C\&R16 catalogue which were not considered. The background colour is the reddening from the \cite{Schlegel1998} dust map with the correction of \cite{Schlafly2010}, clearly highlighting M31 itself. The two rectangle with 'ears' are the observed fields that form the footprint for this study.}
        \label{Fig:footprint}
\end{figure}

The imaging for this study was obtained with MegaCam at the CFHT under programme 22bc04 (PI: Harris). The imaging consists of two fields, seen in Fig. \ref{Fig:footprint} alongside the positions of all GCs from the catalogue of \cite{Caldwell2016} (hereafter C\&R16) and reddening values from the \cite{Schlegel1998} dust map. One field lies further out in the M31 halo and contains EXT8, highlighted in red, alongside a handful of other GCs. The second field lies closer to M31, covering a large portion of the disk and the centre of M31, clearly highlighted by the larger reddening values, encompassing substantially more GCs.

For each field, 5 exposures of 300 seconds in CaHK were obtained to reach a S/N $\gtrsim 40$ for a typical GC (half-light radius of 3 pc and half-light luminosity of $M_V$ = -7, equivalent to $m_V$ $\cong$ 17.5 at M31). Alongside the CaHK exposures, the fields were also observed in the MegaCam u, g and i filters. For each of these three filters, 3 exposures of 30 seconds were obtained per field. The fields were observed on the 25th and 26nd of July 2022 under camera run 22Am07. During both nights, the sky was photometric for the entire observing period. The image quality for each filter remained relatively stable with differences rarely exceeding the pixel scale of MegaCam. The airmass remained stable or decreased slightly across the observing period. The airmass was around $\sim1.1$ on the first night and $\sim1.3$ on the second. The observing logs and main parameters of the observing periods can be found in Table \ref{table:obs_logs}.\\
\\
The GC catalogue presented by C\&R16 offers the largest sample of spectroscopically determined metallicities for M31 GCs within the observational footprint, totalling at 125 GCs. We purposefully stick to spectroscopic metallicities as they are more robust than their photometric counterparts. The C\&R16 catalogue did not include a metallicity for EXT8, so the value of [Fe/H] = -2.91 from \cite{Larsen2020} is assumed. This creates a sample of 126 GCs spanning a metallicity range of -2.9 $\leq$ [Fe/H] $\leq$ +0.4.

Of the 126 GCs only 7 have metallicities below -2. While the sample has a limited amount of very metal-poor GCs, it does contain B157-G212 ([Fe/H] = -2.6 $\pm$ 0.3) and B160-G214 ([Fe/H] = -2.8 $\pm$ 0.4) that are very metal-poor and potentially lie beneath the metallicity floor, making them very valuable for comparison with EXT8.\\
\\
The M31 disk presents a significant challenge in the form of differential extinction. This problem has been tackled by earlier studies of the M31 GC system (e.g \citealt{Barmby2000}), however the uncertainty on these reddening estimates, E(B-V), is often high. The estimated uncertainty on the reddening values in the C\&R16 catalogue is $\pm0.1$ mag in E(B-V). To minimize using these uncertain reddening estimates, we adopt reddening estimates from more robust sources wherever possible. We draw from either colour-magnitude diagrams \citep{Larsen2021, Federici2012} or the \cite{Schlegel1998} dust map, with the correction of \cite{Schlafly2010}, where E(B-V) $\leq$ 0.15. For the remaining GCs, we adopt the reddening values from C\&R16. The C\&R16 catalogue did not possess reddening values for 5 GCs, which were obtained from either \cite{Fan2010} or \cite{Kang2012}. The GCs with the more robust reddening estimates are marked by orange diamonds throughout the figures, while the remaining GCs are marked by blue circles.\\
\\
Values for the extinction coefficients for the MegaCam u, g and i filters have not been published as far as we are aware. It was opted to approximate the extinction coefficients, $c_X = \frac{A_X}{E(B-V)}$, for all filters by considering the filters at their reference wavelength. The \cite{Fitzpatrick1999} extinction law with $R_{V} = 3.1$  was adopted for the approximations, following the results of \cite{Schlafly2011}. The resultant extinction coefficients for the four filters can be found in Table \ref{tab:extinction coefficients}. The extinction coefficient for CaHK was recently calculated by \cite{Martin2023} and only differs by 0.003 from the value adopted here.
\begin{table}
\caption{Approximated extinction coefficients for the MegaCam filters.} 
\label{tab:extinction coefficients}                          
\centering                         
\begin{tabular}{ll}       
\hline\hline                
Filter & $c_{X}$\\    
\hline                       
   CaHK & 4.526\\    
   u & 4.805\\
   g & 3.720\\
   i & 2.003\\ 
\hline                                  
\end{tabular}
\end{table}

\subsection{Photometry} \label{Sec:Phot}
Each of the exposures had come preprocessed by the Elixir pipeline \citep{Magnier2004}, which had debiased, detrended and flatfielded them. The background estimation and aperture photometry was done with SourceExtractor version 2.25.0 \citep{Bertin1996}. For background estimation, a combination of BACK\_SIZE = (36, 36) and BACK\_FILTERSIZE = (8, 8) was used alongside local re-examination around sources (BACKPHOT\_TYPE = LOCAL). The size of the box in which the re-examination took place was set to 13 by 13 arcsecond, BACKPHOT\_THICK = (70, 70), in order to accommodate the size of the largest GCs in our sample.

As it is expected that GCs do not show colour gradients, it was chosen to use a small static aperture for photometry. The size of the aperture was chosen to match the size of the smallest GCs within the sample to prevent the smallest GCs from being heavily influenced by their backgrounds. Using the g-band aperture sizes from \cite{Peacock2010} as a guideline, the diameter of the aperture was chosen to be 3 arcseconds (11.5 pc). This aperture size is larger than the majority of half-light radii reported by \cite{Peacock2010} for the GCs in our sample.

\subsection{Calibration} \label{Sec:Cal}
Calibration of the photometry was done by utilizing synthetic photometry derived from the low-resolution Gaia XP spectra through GaiaXPy\footnote{\url{https://gaia-dpci.github.io/GaiaXPy-website/}} \citep{Gaia2023}. The CaHK filter was already available in GaiaXPy but the MegaCam u, g and i filters were not. As broadband filters needed to be 'standardized', we used the already standardized SDSS system and transformed them to the MegaCam u, g and i magnitudes following the relation defined in the documentation of the MegaCam filter set\footnote{\url{https://www.cadc-ccda.hia-iha.nrc-cnrc.gc.ca/en/megapipe/docs/filt.html}}. The synthetic CaHK magnitudes were not standardized but the sample checked by \cite{Gaia2023} against \textit{Pristine} CaHK magnitudes showed that the synthetic CaHK magnitudes only had an offset of 0.04 mag. The CaHK synthetic magnitudes are in the Vega system, while the transformed SDSS magnitudes are in the AB system. In order to match, the u, g and i magnitudes were transformed to the Vega system by using documented relations\footnote{\url{https://www.cfht.hawaii.edu/Instruments/Imaging/Megacam/specsinformation.html}}.

To get a clean sample of stars with XP spectra from Gaia DR3, a combination of the RUWE and C$^{*}$ diagnostic metrics were used \citep{Lindegren2021, Riello2021}. We adopted the cuts of RUWE < 1.4 and |C$^{*}$| < $\sigma$, where $\sigma$ is the 1$\sigma$ scatter measured as a power-law with magnitude in \cite{Riello2021}. Additionally, we adopted a cut of S/N $\geq$ 30 on the synthetic photometry for u and CaHK, per recommendation of \cite{Gaia2023} to combat the high uncertainty caused by the the low throughput and highly structured nature of the BP band in UV wavelengths. This S/N cut was also applied to the g and i synthetic magnitudes to ensure that only good quality synthetic photometry would be used for calibration. 

After calculating the zeropoints for each of the remaining stars, two cuts were applied to filter out saturated stars and obvious outliers, which are defined as showing $|\text{ZP} - \text{ZP}_{\text{median}}| \geq 0.2$. Using the cleaned sample of calibration stars, the zeropoints for each exposure were calculated as the uncertainty weighted mean. The CaHK zeropoints were additionally corrected for the 0.04 mag offset. The calculated exposure zeropoints were applied to calibrate the magnitude scales of the exposures, after which any obviously erroneous measurements were removed and the remaining measurements for each GC were averaged.

\subsection{Field of view correction} \label{Sec:FoV}
   \begin{figure}
   \centering
   \includegraphics[width=\hsize]{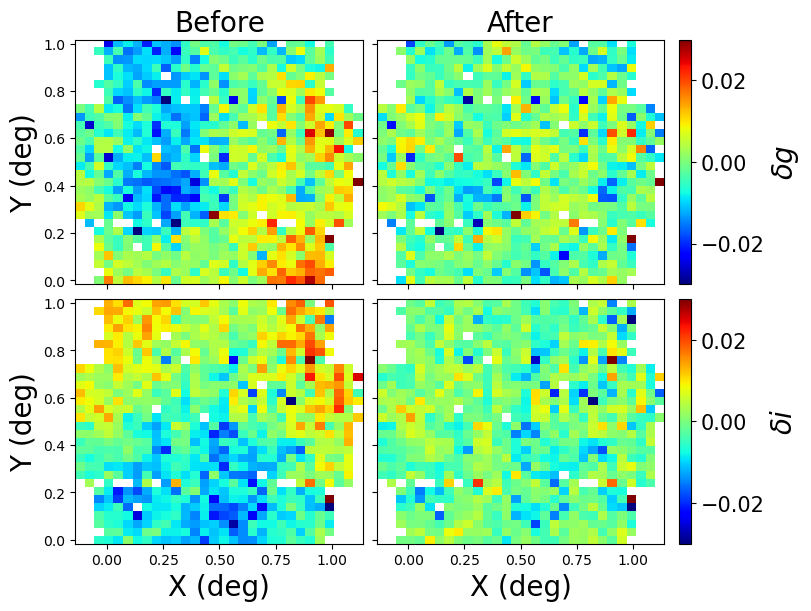}
      \caption{Discretized calibration residuals before and after applying the FoV corrections determined by PhotCalib for the g and i filters. The first column shows the residuals for \textit{g} (top) and \textit{i} (bottom) before correction and the second column shows the residuals after the correction is applied.
              }
         \label{fig:FoVcorrection}
   \end{figure}

It is known that MegaCam exposures suffer from Field of View (FoV) dependent offsets, which are unique in shape and scale for each filter and camera run (e.g. \citealt{Regnault2009,Martin2023}). To determine FoV corrections for our data, a new program called PhotCalib\footnote{\url{https://github.com/zyuan-astro/PhotCalib}} was utilized. PhotCalib is designed for use by \textit{Pristine} to calibrate their data simultaneously for FoV and field dependent effects using machine learning. Simply put, PhotCalib takes uncalibrated magnitudes, positions on the FoV and magnitudes of calibration stars and learns a function dependent on the position on the FoV that minimizes the difference between the uncalibrated and calibration magnitudes. For more details, see \cite{Martin2023}.\\
\\
It has been found that PhotCalib requires about 3000 stars with good quality photometry to produce sufficiently accurate FoV corrections \citep{Martin2023}. Due to the applied quality cuts, this limit cannot be reached for CaHK and u in our data. For the g and i filters, PhotCalib was fed the averaged magnitudes, positions on the FoV and synthetic photometry of the calibration stars. The produced FoV corrections by PhotCalib noticeably improved the flatness of the calibration residuals, as seen in Fig. \ref{fig:FoVcorrection}.

As the FoV corrections are just dependent on filter and camera run, it was possible to use the CaHK corrections obtained by \textit{Pristine} for camera run 22Am07 (Zhen Yuan, private communication). The CaHK corrections for our data were obtained by fitting the corrections from \textit{Pristine} to the CaHK calibration residuals through $\chi^2$ minimization. The majority of corrections are minor, $|\delta CaHK| \leq 0.02$, however for a couple of GCs at the edges of the corrections reached values of $\delta CaHK \sim 0.04$. 
    \begin{figure}
   \centering
   \includegraphics[width=\hsize]{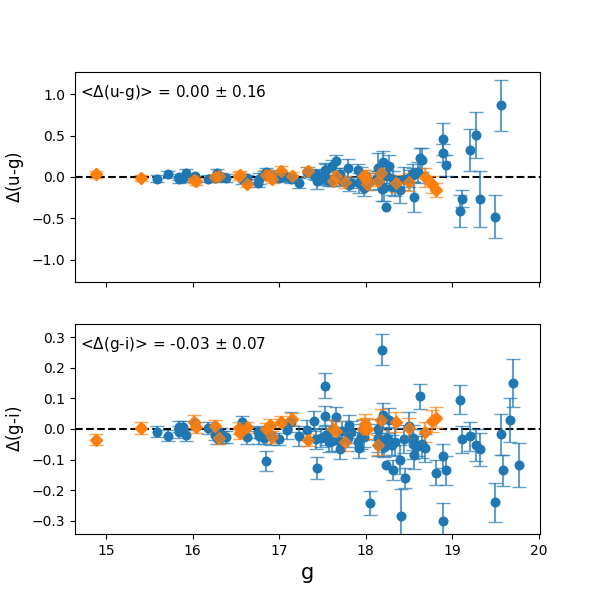}
      \caption{Differences in u-g (top) and g-i (bottom) between our photometry and the \cite{Peacock2010} photometry transformed to MegaCam colours. The text in the upper left of each plot is the average difference and scatter for the entire comparison sample.
              }
         \label{fig:phot_comp}
   \end{figure}
This leaves only the u magnitudes without FoV corrections. The scale of FoV corrections for $u$ is expected to be similar to or slightly larger than that of the CaHK corrections (see Fig. 3 of \citealt{Ibata2017}).\\
\\
The uncertainty on the photometry was initially calculated as the standard deviation from multiple measurements. Where only one measurement was available or the standard deviation was below the average statistical uncertainty, the average statistical uncertainty was adopted instead. 

Comparison with the calibration residuals showed that these calculated uncertainties were insufficient to explain the scatter of the residuals by a constant, the systematic uncertainty. The systematic uncertainty for each filter was determined through a maximum likelihood estimation under the assumption that the calibration residuals followed a Gaussian distribution with $\mu_i = 0$ and $\sigma_i^2 = \sigma_{syn,i}^2 + (\sigma_{phot,i} + \sigma_{sys})^2$, where $\sigma_{syn,i}$ and $\sigma_{phot,i}$ are the uncertainty of the synthetic photometry and the calculated uncertainty for a calibration star and $\sigma_{sys}$ is the to be fitted systematic uncertainty. This yielded: $\sigma_{sys,CaHK}$ = 0.013, $\sigma_{sys,u} = 0.023$, $\sigma_{sys,g} = 0.007$, $\sigma_{sys,i} = 0.006$. For the majority of the GCs, this is the dominant contribution to their uncertainty.

Lastly, the uncertainty regarding background estimation was estimated by performing photometry with a different set of background estimation parameters which produced approximately the same background at larger scales. The average differences between the magnitudes were added to the uncertainties in quadrature with only a small handful of GCs being significantly impacted. The magnitudes and final uncertainties can be found in Table \ref{tab:phot}.\\
\\
Fig. \ref{fig:phot_comp} shows the comparison between our photometry and the published photometry of \cite{Peacock2010}, which is the largest catalogue of \textit{ugriz} photometry for M31 GCs. The \cite{Peacock2010} photometry is SDSS \textit{ugriz} which are transformed to MegaCam \textit{ugriz} using the same transformations as used for calibration. These transformations are largely based on stars, so applying these to GCs can introduce more scatter or systematic errors. Additionally the shown uncertainties do not include the uncertainty of the transformations, so it is expected that the scatter is larger than the plotted uncertainties. However, it can still be used to check for consistency between the sets of photometry.
  \begin{figure*}
   \resizebox{\hsize}{!}
            {\includegraphics[height=6mm]{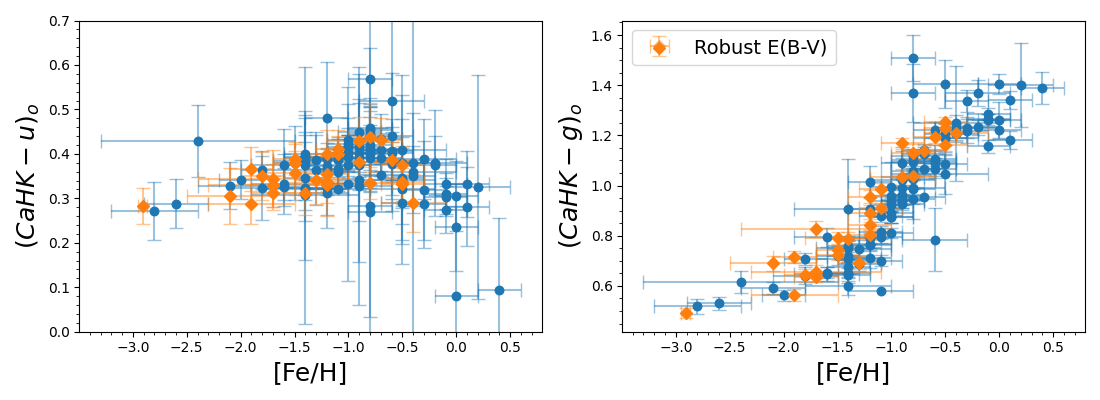}}
      \caption{(CaHK-u)$_{o}$ (left) and (CaHK-g)$_{o}$ (right) plotted versus metallicity for the GC sample. The GCs with robust reddening estimates are marked with orange diamonds and the remaining are marked by blue circles. EXT8 can be seen separating itself from the other metal-poor GCs as the left most GC in each plot. Note the difference in vertical scale between the two panels (0.1 vs 0.2 mag major tick spacing).
              }
         \label{fig:CaHKindices}
   \end{figure*}
   
There is good consistency for u-g, but there is a 0.03 mag offset for g-i. Lacking correlations with GC structural parameters or the SDSS colours, the small offset is likely caused by the transformations and is not an actual offset in the photometry. There is, however, a group of outliers in g-i. Visual inspection of these GCs revealed that they lie in crowded or dense regions, mostly near the core of M31, making background estimation very difficult. This means that the outliers are caused by the different choices in background estimation. The offset of these outliers may be amplified by the transformations, so we believe that there is consistency between the sets of photometry.

\section{Results}
With the express goal of CaHK photometry being to measure the strength of the Ca {\sc ii} H \& K lines, the combinations of CaHK with the u and g filters are of particular interest as these filters directly flank CaHK on the shorter and longer wavelengths respectively. This allows the u and g magnitudes to act as a measurement of the local continuum which normalises the CaHK magnitude. Therefore, we investigate CaHK-u and CaHK-g.\\
\\
Fig. \ref{fig:CaHKindices} shows each of the CaHK colours as a function of metallicity for the GC sample. Despite earlier concerns with regards to the quality of the reddening estimates, the similarity seen between GCs with and without robust reddening estimates shows that this is not an issue at the scale of the colours.

In the left plot of Fig. \ref{fig:CaHKindices}, it can be seen that CaHK-u shows a small range of values, $\Delta(CaHK-u) \sim 0.2$, across the entire metallicity range and suggests a non-monotonic relation by considering GCs with [Fe/H] $\geq$ -0.8 separately from GCs with [Fe/H] $\leq$ -0.8. $(CaHK-u)_{o}$ shows a negative correlation with metallicity for [Fe/H] $\geq$ -0.8, as $(CaHk-u)_{o}$ decreases in value from $\sim$0.4 at [Fe/H] = -0.8 to $\sim$0.27 at [Fe/H] = 0 with a suggestion of even lower values for super-solar metallicities. A Spearman correlation test gives a correlation coefficient of $\rho_{\text{[Fe/H]} \geq -0.8} = -0.67^{+0.18}_{-0.13}$, where the uncertainty is the 95 percent confidence interval. For [Fe/H] $\leq$ -0.8, $(CaHK-u)_{o}$ shows a positive correlation as its value decrease from  $\sim$0.4 back to a similar value of the solar metallicity GCs for the most metal-poor GCs as the metallicity decreases with $\rho_{\text{[Fe/H]} \leq -0.8} = 0.52^{+0.18}_{-0.14}$. The non-monotonic relation with metallicity makes the interpretation of $(CaHK-u)_o$ as a measure of CaHK strength difficult as a single $(CaHK-u)_0$ value can correspond to two, potentially drastically different, metallicities. Therefore $(CaHK-u)_o$ must always be used in conjunction with another colour that can break the degeneracy.

In the right plot of Fig. \ref{fig:CaHKindices}, $(CaHK-g)_{o}$ shows a strong positive correlation with metallicity ($\rho_{CaHK-g} = 0.92^{+0.03}_{-0.02}$), changing about 0.8 magnitudes across the entire metallicity range, going from $\sim$1.3 at solar metallicity down to $\sim$0.5 at the very metal-poor end. $(CaHK-g)_{o}$ seems to experience a break around [Fe/H] = -1.4, as the slope of the relation noticeably decreases. For [Fe/H] $\leq$ -1.4, the correlation coefficient decreases to $\rho_{\text{[Fe/H]} \leq -1.4} = 0.60^{+0.18}_{-0.28}$ while for [Fe/H] $\geq$ -1.4 the correlation coefficient remains relatively similar $\rho_{\text{[Fe/H]} \geq -1.4} = 0.89^{+0.03}_{-0.05}$. The relation shown by $(CaHK-g)_o$ is reminiscent of the bilinear relation commonly found for metallicity sensitive spectral indices for GCs (e.g. \citealt{Burstein1984}). This suggests that $(CaHK-g)_o$ provides a simple measurement of the strength of the Ca {\sc ii} H \& K lines and makes the interpretation of $(CaHK-g)_o$ straightforward with low $(CaHK-g)_o$ values corresponding with weak Ca {\sc ii} H \& K lines/low metallicity and vice versa.\\
\\
For our purpose, the interest lies in the metal-poor regime ([Fe/H] $\leq$ -1.5) with the main question to answer being if EXT8 can be separated from regular metal-poor GCs. EXT8, as the most metal-poor GC in the sample, can be easily identified as the left most GC in each of the plots in Fig. \ref{fig:CaHKindices} as an orange diamond. In $(CaHK-g)_o$, EXT8 clearly stands out from the other metal-poor GCs as it has the lowest $(CaHK-g)_o$ value, $(CaHK-g)_o = 0.49\pm0.02$, by which it is separated from the closest regular metal-poor GCs by 0.07. In $(CaHK-u)_o$, EXT8 does not separate itself as clearly from the other metal-poor GCs as its $(CaHK-u)_o$ value, $(CaHK-u)_o = 0.28\pm0.04$, is only separated by 0.004 from the nearest regular metal-poor GC.

Alongside EXT8, the two potential GCs below the metallicity floor in the sample, B160-G214 ([Fe/H] = -2.8 $\pm$ 0.4) and B157-G212 ([Fe/H] = -2.6 $\pm$ 0.3), show similar behaviour as EXT8. In $(CaHK-g)_o$, they separate themselves from the more metal-rich GCs, but show a slightly higher $(CaHK-g)_o$ than EXT8, hinting at a good metallicity sensitivity of $(CaHK-g)_o$ even near the extremely metal-poor regime. In $(CaHK-u)_o$, they show equivalent values to EXT8 within their uncertainty with B160-G214 showing a slightly lower value.

Considering all metal-poor GCs ([Fe/H] $\leq$ -1.5) suggests that the behaviour of both CaHK colours in the metal-poor regime is linear, although the sparse amount of GCs in this regime makes it difficult to infer anything but linear behaviour. Regardless, fitting a linear relation to the metal-poor GCs through unweighted least-squares gives the following relations:
\begin{equation}
\begin{split}
    (CaHK-u)_{o} &= (0.064\pm0.012)\text{[Fe/H]} + (0.456\pm0.023)\\
    (CaHK-g)_{o} &= (0.177\pm0.032)\text{[Fe/H]} + (0.994\pm0.062)
\end{split}
\end{equation}
The RMS of the fits are 0.02 mag and 0.06 mag for $(CaHK-u)_o$ and $(CaHK-g)_o$ respectively. Despite the substantial difference in slope, for both $(CaHK-u)_o$ and $(CaHK-g)_o$ the uncertainty on metallicity estimates suggested by the RMS would be 0.3 dex at a 1$\sigma$ level.
\subsection{Comparison with other GC colours}
EXT8 and the other potential GCs below the metallicity floor have shown that CaHK photometry holds up to its suggested potential for candidate selection with $(CaHK-g)_o$ providing a clean selection of these GCs, but how do the CaHK colours compare to potential alternative GC colours that could be considered for the same purpose?

One might consider using u-g as the u filter covers the metallicity-sensitive shorter wavelengths and requires a lower integration time to reach similar S/N due to its broadband nature compared to CaHK. From literature, the g-z colour from SDSS filters, or equivalently F450W-F850L from HST filters, has seen significant usage in studying extragalactic GC systems alongside spectroscopic campaigns (e.g. \citealt{Peng2006,Sinnott2010,Usher2012,Villaume2019,Fahrion2020}). This makes g-z an appealing option as it is more readily available and has been spectroscopically calibrated.

\begin{figure}
   \centering
   \includegraphics[width=\hsize]{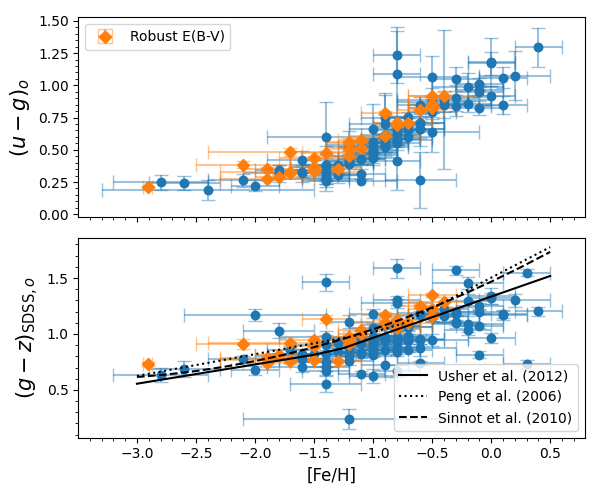}
      \caption{The broadband colours, $(u-g)_o$ (top) and $(g-z)_{\mathrm{SDSS},o}$ (bottom), plotted against
metallicity for our GC sample. For $(g-z)_{\mathrm{SDSS},o}$, the spectroscopically calibrated CMRs from \cite{Peng2006} (dotted), \cite{Sinnott2010} (dashed) and \cite{Usher2012} (solid) are also plotted. For both colours, the GCs below the metallicity do not separate clearly from other GCs.
              }
         \label{fig:broadband_comparison}
   \end{figure}

Fig. \ref{fig:broadband_comparison} shows $(u-g)_o$ in the top panel and $(g-z)_{\mathrm{SDSS},o}$ from \cite{Peacock2010} in the lower panel. The lower panel with $(g-z)_{\mathrm{SDSS},o}$ also shows the spectroscopically determined colour-metallicity relations (CMRs) from \cite{Peng2006}, \cite{Sinnott2010} and \cite{Usher2012}. The extinction coefficients for SDSS g and z filters were approximated in the same way as for the MegaCam filters, yielding $c_{g,\mathrm{SDSS}} = 3.751$ and $c_{z,\mathrm{SDSS}} = 1.587$.
   
From Fig. \ref{fig:broadband_comparison}, it can be seen that the GCs below the metallicity floor do not separate as well in both colours as they do in $(CaHK-g)_o$. In the GC sample the colour values of the GCs below the metallicity floor can also be found for GCs up to [Fe/H] = -2 for $(u-g)_o$ and for $(g-z)_{\mathrm{SDSS},o}$ all the way up to [Fe/H] = -0.6. For $(g-z)_{\mathrm{SDSS},o}$, the majority of these GCs have the more uncertain reddening estimate but even for the GCs with the more robust reddening EXT8 has a similar $(g-z)_{\mathrm{SDSS},o}$ value to GCs up to [Fe/H] = -1.3. 

All three spectroscopically calibrated CMRs for $(g-z)_{\mathrm{SDSS},o}$ fit reasonably well to the data and predict the blue colour of B157 and B160 within their uncertainty. EXT8 is however slightly redder and would be missed when utilizing any of the three CMRs. This suggests that the metallicity sensitivity of $(g-z)_{\mathrm{SDSS},o}$ is lower than predicted by extrapolation of the literature CMRs.\\
\\
Performing the same linear fits for $(u-g)_o$ and $(g-z)_{\mathrm{SDSS},o}$ as for the CaHK colours yields the following relations: 
\begin{equation}
\begin{split}
    (u-g)_{o} &= (0.111\pm0.029)\text{[Fe/H]} + (0.535\pm0.056)\\
    (g-z)_{\mathrm{SDSS},o} &= (0.118\pm0.047)\text{[Fe/H]} + (1.022\pm0.090)
\end{split}
\end{equation}
The RMS of the fits are 0.05 mag and 0.08 mag for $(u-g)_o$ and $(g-z)_{\mathrm{SDSS},o}$ respectively. The RMS of the fits suggest an uncertainty of 0.5 dex and 0.7 dex on metallicity estimates using $(u-g)_o$ and $(g-z)_{\mathrm{SDSS},o}$. Both of these are worse than the metallicity uncertainty of the CaHK colours. 

The literature $(g-z)_{\mathrm{SDSS},o}$ CMRs report uncertainties on metallicity estimates between 0.17 and 0.3 dex, which would make them better or on par with the CaHK colours. The difference between the reported uncertainties and those found here has to do with the different GCs, data and the metallicity range over which was fitted. The high uncertainty on reddening estimates can play a significant role for $(g-z)_{\mathrm{SDSS},o}$, however the GCs with more robust reddening estimates show a similar RMS.\\
\\
The lower uncertainty on metallicity estimations mean that the CaHK colours provide a less contaminated selection compared to $(u-g)_o$ and $(g-z)_{\mathrm{SDSS},o}$. Using the GC metallicity distributions of the Galaxy \citep[2010 edition]{Harris1996} and M31 (C\&R16) with the fit RMS as a Gaussian scatter on the CMRs, the CaHK colours have a false positive identification rate around 4 percent for GCs with -2.5 $\leq$ [Fe/H] $\leq$ -1.5 to be identified as a GC with [Fe/H] $\leq$ -2.5. This is a factor 2 better than $(u-g)_o$ and a factor 3.8 better than $(g-z)_{\mathrm{SDSS},o}$. The high-S/N and high-resolution of \cite{Larsen2020} for EXT8, a relatively close by and bright GC, already had an integration time of 40 minutes. This means that reducing the selection contamination by a factor of 2 is a significant reduction in telescope time for future spectroscopic observations. This is especially true for more distant systems and massive galaxies that can host several hundreds up to thousands of GCs with [Fe/H] $\leq$ -1.5 (e.g. \citealt{Harris2023}).

A PCA test was done with the available colours to see if any colour combination would perform better than the CaHK colours alone. The best PCA component, predominantly relying on $(CaHK-g)_o$ and $(u-g)_o$, showed an uncertainty on metallicity estimation of 0.4 dex, slightly worse than the individual CaHK colours.

\subsection{Horizontal branch morphology sensitivity}
Metallicity is one of the main parameters that determines horizontal branch (HB) morphology with metal-rich GCs showing cooler red HB populations while metal-poor GCs show hotter blue HB populations (e.g. \citealt{Arp1952}). The Galactic GCs (GGCs) at similar metallicity, however, show a wider range of morphologies with some showing both a red and a blue HB or an extended HB with a characteristic tail. This phenomenon is known as the 'second parameter' problem (e.g. \citealt{VandenBergh1965,Graton2010,Dotter2010}). 

The differences in HB morphology have been tied to other GC properties. The main culprits are GC age and helium variations within a GC (e.g. \citealt{Sandage1967,Searle1978,Lee1994,Milone2014}). A younger age causes the HB to be redder at the same metallicity and larger helium variations cause a more extended HB. The latter is tied to the multiple population phenomenon which can additionally explain the existence of bimodal HB morphologies (e.g. \citealt{Marino2011}).

It is well known that the HB morphology significantly influences the colours and line indices of GCs, especially the Balmer lines (e.g. \citealt{Lee2000,Lee2002,Schiavon2004,Percival2011}). The CaHK colours are likely not an exception to this as the Ca {\sc ii} H \& K lines are located in the UV and Ca {\sc ii} H is blended with the H$_\epsilon$ Balmer line. If the CaHK colours are sufficiently sensitive to HB morphology, it can cause complications for candidate selection. An example would be if an intermediate metallicity GC with a very blue HB morphology mimics the CaHK colours of a GC at or below the metallicity floor.\\

\begin{figure}
   \centering
   \includegraphics[width=\hsize]{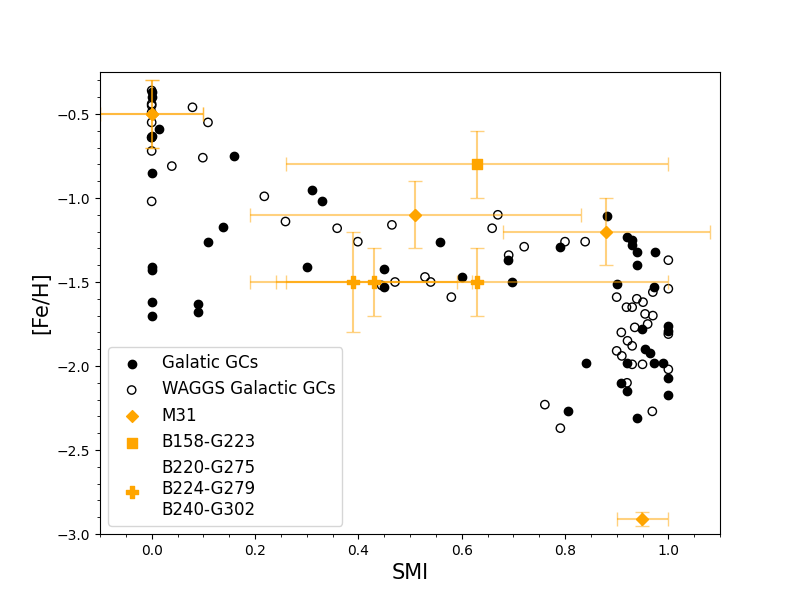}
      \caption{A SMI versus metallicity diagram. The 9 M31 GCs are plotted in orange. The GGCs are plotted as black circles. The SMI for the GGCs is a combination of data from \cite{Perina2012} and \cite{Lee1994}. The \cite{Lee1994} horizontal branch ratio (HBR) values were transformed to SMI by the relation of \cite{Preston1991}. The Galactic GC metallicities were taken from \cite[2010 edition]{Harris1996} The GGCs that have WAGGS spectra \citep{Usher2017} are marked by open black circles.
              }
         \label{fig:HBMorph_showcase}
\end{figure}
   
\begin{figure}
   \centering
   \includegraphics[width=\hsize]{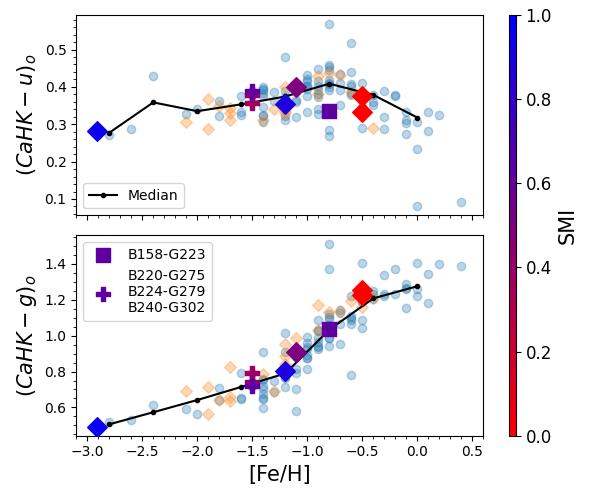}
      \caption{The CaHK CMRs of Fig. \ref{fig:CaHKindices} overlaid with the HB morphology measurements from \cite{Perina2012} and \cite{Larsen2021} as a colour. A red colour corresponds to a red branch and blue corresponds a blue branch. The black line shows the median trend for each of the CaHk colours with metallicity calculated with a binwidth of 0.4 dex. Note the difference in vertical scale between the two panels (0.1 vs 0.2 mag major tick spacing).
              }
         \label{fig:HBMorph}
\end{figure}

\noindent There are only a small number of HB morphology measurements for GCs in the our sample, 8 measurements from \cite{Perina2012} and one for EXT8 from \cite{Larsen2021}. \cite{Perina2012} measured the HB morphology with the Simplified Mironov Index, SMI $\equiv \frac{B}{B+R}$, where $B$ and $R$ are the number of HB stars bluer or redder than a given colour threshold, usually chosen in the middle of the RR Lyrae instability strip. Small SMI values indicate a red dominant branch while large SMI values indicate a blue dominant branch. All but one (B407-G352; SMI = 0 $\pm$ 0.1) of the SMI values from \cite{Perina2012} for our sample were derived from lesser quality CMDs and are considered to be lower limits as blue HB stars could fall below the limiting magnitude of the CMDs. 

Fig. \ref{fig:HBMorph_showcase} shows the HB morphologies for the 9 M31 GCs alongside that of the GGCs. It can be seen that B158-G223 (square) shows a substantially higher SMI (0.63 $\pm$ 0.37) for its metallicity, suggesting the presence of an unusually large blue HB component. The set of B220-G275, B224-G279 and B240-G302 (pluses) show redder branches (SMI = 0.39 $\pm$ 0.2, 0.43 $\pm$ 0.19 and 0.63 $\pm$ 0.37), similar to M3 among the Galatic GCs (e.g. \citealt{Rey2001}), meaning these GCs are relatively younger.

Fig. \ref{fig:HBMorph} shows the CaHK CMRs with the SMI values of the 9 GCs plotted overtop as a colour alongside the median trend for the CMRs. Only B158-G223 in $(CaHK-u)_o$ seems to strongly deviate from the median trend by 0.07 mag. B158-G223 does, however, lie at the edge of the inner field covering the disk of M31. This suggests that it is subject to a large FoV correction for its u magnitude, which would shift its $(CaHK-u)_o$ value towards the median. 

In addition to the 9 M31 GCs, we investigate the GGCs by generating synthetic colours from the WAGGS spectra \citep{Usher2017}. The WAGGS spectra are a set of relative calibrated spectra covering 62 GGCs (marked by open circles in Fig. \ref{fig:HBMorph_showcase}) from 3270 \AA\space to 9050 \AA. 

Similar to the M31 GCs, we find that differing HB morphologies at the same metallicity do not strongly affect the CaHK colours. It should be kept in mind that the WAGGS spectra, like other IL spectra, only capture a fraction of the total cluster light (the median fraction of V-band light is 0.12; \citealt{Usher2017}) making them susceptible to stochastic variations and repeat observations showed that the UV wavelengths of the spectra have variations up to 12 percent. 

In the end, we do not find any indication of a strong HB morphology sensitivity such that differing morphologies might present an issue for candidate selection. However, neither of the datasets contained metal-poor, red HB GCs ([Fe/H] $\leq$ -1.5 \& SMI $\leq$ 0.2), of which more extreme versions of the GGCs were recently discovered in the outer halo of M31 \citep{McGill2025}. Ultimately, this assessment is limited by the amount and quality of the available data and should be followed up with a more in-depth assessment with more robust data.

\section{Conclusion}
In order to lay the groundwork for a more organized search for other GCs like EXT8, we investigated the behaviour of the CaHK colours, CaHK-u and CaHK-g, with metallicity for the purpose of candidate selection with a sample of M31 GCs, including EXT8.

$(CaHK-u)_o$ shows a non-monotonic CMR with a maximum around [Fe/H] = -0.8, which results in a small colour range and that it must be used in conjunction with another colour to break the degeneracy. $(CaHK-g)_o$ shows an approximately bilinear CMR, similar to that of metallicity sensitive spectral indices, suggesting that $(CaHK-g)_o$ is a straightforward measurement of the Ca {\sc ii} H \& K lines. Despite significant differences in metallicity sensitivity in the metal-poor regime, with $(CaHK-g)_o$ showing 0.177 $\pm$ 0.032 mag/dex, the RMS of the relations for both CaHK colours suggest an uncertainty of 0.3 dex on metallicity estimates from the colours. In comparison with the broadband colours $(u-g)_o$ and $(g-z)_{\mathrm{SDSS},o}$, the CaHK colours could reduce false positive identification rates and by extension the follow-up observation time by at least a factor 2. We also investigated the influence of the HB morphology based on available data, and find no indication that HB morphology affects the CaHK colours to a degree that would pose a problem for candidate selection.

The results presented here are just an initial test for CaHK photometry for GCs and far from conclusive, given the unavoidable sparseness of GCs in the metal-poor regime, limited robust information of M31 GC properties in our sample and potential differences between GC systems (e.g. \citealt{Usher2015,Powalka2016}). Despite these limitations, we find that CaHK photometry for GCs does hold up to it suggested potential and can be used as an effective indicator of metallicity in order to identify candidates for massive GCs below the metallicity floor.

While $(CaHK-g)_o$ can in principle be used alone for candidate selection, using it in conjunction with other colours naturally provides more information for candidate selection. Primarily, the inclusion of $u$-based colours would fully characterize the wavelength region around the Ca {\sc ii} H \& K lines. The fitted relations for all colours shown in this study suggest that GCs below the metallicity floor should have the following colour values: $(CaHK-g)_o \leq 0.551$; $(CaHK-u)_o \leq 0.295$; $(u-g)_o \leq 0.258$ and $(g-z)_{SDSS,o} \leq 0.727$.\\
\\
This study marks the beginning of the search for massive GCs below the metallicity floor in order to learn if and/or how this previously unknown class of GC fits into our understanding of GC formation. With the expectation that, in general, GCs below the metallicity floor must originate from low-mass/dwarf galaxies, the haloes of massive galaxies make for a good starting point as these haloes have accreted a substantial amount of dwarf systems. 

The best estimates of the rarity of EXT8-like GCs are 0.05 percent (emperical; \citealt{Beasely2019}) and 0.03 percent (simulation; \citealt{Usher2018}) which means that even in massive haloes that can hold up to several thousands of GCs (e.g. \citealt{Durrell2014,Harris2023}), only a handful of EXT8-like GCs are expected. This means the effectiveness of the CaHK colours will be key for the search effort. 

\begin{acknowledgements}
The authors thank the referee for their comments and suggestions that helped improve the quality of this manuscript. BvH is grateful to Zhen Yuan and Nicolas Martin for kindly sharing the CaHK FoV correction for camera run 22Am07. ES warmly acknowledges helpful discussions with Eric Peng on this subject.\\
WEH acknowledges support from the Natural Sciences and Engineering Research Council of Canada. ES acknowledges funding through VIDI grant "Pushing Galactic Archeology to its limits" (with project number VI. Vidi.193.093) which is funded by the Dutch Research Council (NWO). This research has been partially funded from a Spinoza award by NWO (SPI 78-411).\\
This work has made use of \textsc{NUMPY} \citep{VandeWalt2011}, \textsc{MATPLOTLIB} \citep{Hunter2007} and \textsc{SCIPY} \citep{2020SciPy}.
\end{acknowledgements}

\bibliographystyle{aa}
\bibliography{references}

\begin{appendix}

\begin{table*}[h!]
\section{Observing logs}
\caption {The observing logs for each of the fields and the main parameters for each exposure obtained.}
\label{table:obs_logs} 
\centering
\begin{tabular}{c c|c c c c} 
\hline\hline            
Sky (J2000) & Date (UT) & Filter & Exposure Time (s) & Image Quality (\arcsec) & Airmass\\ 
\hline
  \multicolumn{6}{l}{\it Outer field}\\
\hline
    RA: 0:51:20.40 & 14:07 - 14:51 & CaHK & 5$\times$300 & [0.91,0.80,0.83,0.86,0.90] & [1.100,1.095,1.095,1.087,1.084]\\
    DEC: 41:30:00.0& 25/7/2022 &u& 3$\times$32 & [0.94,0.96,0.72] & [1.081,1.081,1.080]\\
   &&g& 3$\times$30 & [0.67,0.67,0.69] & [1.079,1.079,1.079]\\
   &&i& 3$\times$30 & [0.76,0.66,0.62] & [1.078,1.078,1.078]\\
   \hline
     \multicolumn{6}{l}{\it Inner field}\\
\hline
   RA: 0:45:44.40 & 11:39 - 12:23 & CaHK & 5$\times$300 & [0.97,0.76,0.69,0.73,0.75] & [1.440,1.413,1.388,1.365,1.343]\\
    DEC: 41:30:00.0& 26/7/2022 &u& 3$\times$32 & [0.83,0.80,0.75] & [1.316,1.312,1.308]\\
   &&g& 3$\times$30 & [0.69,0.75,0.70] & [1.298,1.294,1.290]\\
   &&i& 3$\times$30 & [0.59,0.66,0.57] & [1.281,1.277,1.274]\\
   \hline
\end{tabular}
\end{table*}

\onecolumn
\section{Globular cluster parameters \& photometry}
\begin{longtable}{lllllllll}
\caption{Metallicity, aperture photometry results and reddening for the sample of M31 GCs. The number in the parentheses with the magnitudes is the uncertainty (1$\sigma$) on the magnitudes.}\\
\hline
\hline
Name &  [Fe/H] & Source &  $CaHK$ &     $u$ &     $g$ &     $i$ &  E(B-V) & Source \\
\hline
\endfirsthead
\caption{Continued.}\\
\hline
Name &  [Fe/H] & Source &  $CaHK$ &     $u$ &     $g$ &     $i$ &  E(B-V) & Source \\
\hline
\endhead
\hline
\endfoot
\hline
\endlastfoot
     B337D & -1.7 $\pm$ 0.7 &           C11 & 19.65(0.03) & 19.32(0.06) & 18.77(0.01) & 17.36(0.02) &    0.06 &           SFD \\
 B403-G348 & -0.9 $\pm$ 0.2 &           C11 & 18.26(0.02) & 17.90(0.04) & 17.02(0.01) & 15.44(0.01) &    0.09 &           SFD \\
 B405-G351 & -1.2 $\pm$ 0.2 &           C11 & 16.89(0.02) & 16.55(0.04) & 16.02(0.01) & 14.63(0.01) &    0.08 &           F12 \\
 B407-G352 & -0.5 $\pm$ 0.2 &           C\&R16 & 18.20(0.02) & 17.86(0.04) & 16.89(0.01) & 15.26(0.01) &    0.10 &        F12 \\
      EXT8 & -2.91 $\pm$ 0.04 &         L20 & 16.85(0.02) & 16.59(0.04) & 16.31(0.01) & 15.11(0.01) &    0.06 &           L21 \\
      NB16 & -1.4 $\pm$ 0.5 &           C11 & 20.58(0.16) & 20.47(0.24) & 19.13(0.12) & 16.98(0.05) &    0.68 &           C11 \\
      B260 & -0.6 $\pm$ 0.3 &           C11 & 21.36(0.12) & 21.12(0.22) & 19.77(0.03) & 17.04(0.02) &    1.00 &           C11 \\
 B116-G178 & -0.6 $\pm$ 0.2 &           C11 & 19.44(0.03) & 19.24(0.06) & 17.80(0.01) & 15.37(0.01) &    0.72 &           C11 \\
 B119-NB14 & -0.8 $\pm$ 0.3 &           C11 & 19.21(0.04) & 18.86(0.08) & 18.05(0.02) & 16.45(0.02) &    0.21 &           C11 \\
     B068D & -0.5 $\pm$ 0.4 &           C11 & 20.25(0.08) & 19.87(0.15) & 19.14(0.02) & 17.44(0.03) &    0.08 &           C11 \\
 B122-G181 & -1.2 $\pm$ 0.2 &           C11 & 20.46(0.06) & 20.20(0.11) & 18.81(0.02) & 16.33(0.01) &    0.79 &           C11 \\
     B072D & -0.9 $\pm$ 0.3 &           C11 & 20.79(0.11) & 20.51(0.24) & 19.63(0.02) & 18.08(0.02) &    0.16 &           C11 \\
      B129 & -0.8 $\pm$ 0.2 &           C11 & 20.82(0.11) & 20.89(0.34) & 18.46(0.01) & 15.28(0.01) &    1.24 &           C11 \\
      B262 & -1.3 $\pm$ 0.3 &           C11 & 19.00(0.03) & 18.66(0.07) & 18.21(0.01) & 16.77(0.01) &    0.06 &           C11 \\
 B132-NB15 & -0.5 $\pm$ 0.3 &           C11 & 20.05(0.09) & 19.79(0.17) & 18.41(0.02) & 16.55(0.08) &    0.30 &           C11 \\
     B078D & -0.4 $\pm$ 0.4 &           C11 & 21.64(0.23) & 21.42(0.54) & 20.00(0.03) & 17.89(0.02) &    0.49 &           C11 \\
 B135-G192 & -1.8 $\pm$ 0.2 &           C11 & 17.69(0.02) & 17.40(0.04) & 16.75(0.01) & 15.12(0.01) &    0.28 &           C11 \\
 B264-NB19 & -2.4 $\pm$ 0.9 &           C11 & 19.07(0.04) & 18.70(0.07) & 18.28(0.02) & 16.74(0.05) &    0.21 &           C11 \\
 B137-G195 & -1.5 $\pm$ 0.2 &           C11 & 19.82(0.03) & 19.59(0.08) & 18.68(0.01) & 16.67(0.01) &    0.52 &           C11 \\
      B138 &  0.0 $\pm$ 0.2 &           C11 & 18.93(0.03) & 18.68(0.07) & 17.53(0.02) & 15.79(0.02) &    0.22 &           C11 \\
     AU010 & -0.5 $\pm$ 0.3 &           C11 & 19.55(0.05) & 19.29(0.11) & 18.19(0.03) & 16.42(0.01) &    0.22 &           C11 \\
 B141-G197 & -1.4 $\pm$ 0.2 &           C11 & 18.72(0.02) & 18.43(0.05) & 17.70(0.01) & 16.03(0.01) &    0.32 &           C11 \\
 B143-G198 & -0.1 $\pm$ 0.2 &           C11 & 18.19(0.02) & 17.95(0.05) & 16.76(0.01) & 15.05(0.01) &    0.34 &           C11 \\
      B144 &  0.1 $\pm$ 0.2 &           C11 & 18.82(0.03) & 18.55(0.07) & 17.44(0.01) & 15.76(0.02) &    0.24 &           C11 \\
     B090D & -0.3 $\pm$ 0.2 &           C11 & 19.36(0.04) & 19.11(0.09) & 17.92(0.01) & 16.10(0.01) &    0.13 &           C11 \\
 B147-G199 & -0.1 $\pm$ 0.2 &           C11 & 17.97(0.02) & 17.73(0.05) & 16.58(0.01) & 14.87(0.01) &    0.13 &           C11 \\
      B266 & -1.0 $\pm$ 0.2 &           C11 & 20.62(0.10) & 20.43(0.19) & 19.21(0.01) & 17.12(0.01) &    0.52 &           C11 \\
 B148-G200 & -1.1 $\pm$ 0.2 &           C11 & 17.31(0.02) & 17.00(0.04) & 16.39(0.01) & 14.92(0.01) &    0.28 &           C11 \\
      B269 & -0.8 $\pm$ 0.2 &           C11 & 21.06(0.11) & 20.64(0.23) & 19.67(0.02) & 17.46(0.02) &    0.51 &           C11 \\
 B151-G205 & -0.6 $\pm$ 0.2 &           C11 & 17.23(0.02) & 16.99(0.04) & 15.71(0.01) & 13.72(0.01) &    0.53 &           C11 \\
    SK059A & -0.2 $\pm$ 0.3 &           C\&R16 & 19.95(0.05) & 19.63(0.11) & 18.40(0.01) & 16.52(0.01) &    0.22 &           K12 \\
 B152-G207 & -0.7 $\pm$ 0.2 &           C11 & 18.06(0.02) & 17.70(0.04) & 16.85(0.01) & 15.29(0.01) &    0.18 &           C11 \\
 B356-G206 & -1.4 $\pm$ 0.2 &           C11 & 18.64(0.02) & 18.36(0.04) & 17.76(0.01) & 16.20(0.01) &    0.11 &           SFD \\
 B154-G208 & -0.2 $\pm$ 0.2 &           C11 & 18.90(0.03) & 18.58(0.06) & 17.54(0.01) & 15.89(0.01) &    0.17 &           C11 \\
 B155-G210 & -0.5 $\pm$ 0.2 &           C11 & 19.84(0.04) & 19.60(0.08) & 18.60(0.01) & 16.89(0.01) &    0.19 &           C11 \\
 B156-G211 & -1.2 $\pm$ 0.2 &           C11 & 18.44(0.02) & 18.16(0.05) & 17.61(0.01) & 16.22(0.01) &    0.09 &           C11 \\
 B157-G212 & -2.6 $\pm$ 0.3 &           C11 & 18.75(0.02) & 18.48(0.05) & 18.14(0.01) & 16.90(0.02) &    0.09 &           C11 \\
 B158-G213 & -0.8 $\pm$ 0.2 &           C11 & 16.52(0.02) & 16.21(0.03) & 15.41(0.01) & 13.85(0.01) &    0.09 &           F12 \\
 B160-G214 & -2.8 $\pm$ 0.4 &           C11 & 19.14(0.03) & 18.89(0.06) & 18.55(0.01) & 17.32(0.01) &    0.09 &           C11 \\
 B161-G215 & -1.1 $\pm$ 0.2 &           C11 & 17.85(0.02) & 17.53(0.04) & 16.91(0.01) & 15.47(0.01) &    0.18 &           C11 \\
    SK122C & -0.8 $\pm$ 0.2 &           C\&R16 & 21.29(0.09) & 21.10(0.22) & 19.57(0.02) & 17.28(0.01) &    0.27 &           K12 \\
 B162-G216 & -0.5 $\pm$ 0.2 &           C11 & 19.64(0.04) & 19.32(0.09) & 18.27(0.01) & 16.51(0.01) &    0.21 &           C11 \\
 B163-G217 & -0.1 $\pm$ 0.2 &           C11 & 17.36(0.02) & 17.11(0.04) & 15.92(0.01) & 14.17(0.01) &    0.21 &           C11 \\
 B164-V253 & -0.3 $\pm$ 0.3 &           C11 & 19.86(0.04) & 19.54(0.08) & 18.45(0.01) & 16.69(0.01) &    0.23 &           C11 \\
 B165-G218 & -2.0 $\pm$ 0.2 &           C11 & 17.82(0.02) & 17.52(0.04) & 17.13(0.01) & 15.82(0.01) &    0.15 &           C11 \\
      B167 & -0.4 $\pm$ 0.2 &           C11 & 19.29(0.03) & 18.99(0.06) & 17.94(0.01) & 16.23(0.01) &    0.18 &           C11 \\
      B168 & -0.6 $\pm$ 0.2 &           C11 & 20.61(0.06) & 20.38(0.13) & 18.90(0.01) & 16.44(0.01) &    0.76 &           C11 \\
 B272-V294 & -0.9 $\pm$ 0.2 &           C11 & 20.78(0.06) & 20.57(0.17) & 19.32(0.01) & 17.30(0.01) &    0.46 &           C11 \\
 B172-G223 & -0.6 $\pm$ 0.2 &           C11 & 18.67(0.03) & 18.31(0.05) & 17.42(0.01) & 15.80(0.01) &    0.17 &           C11 \\
 B173-G224 & -0.8 $\pm$ 0.2 &           C11 & 19.50(0.04) & 19.11(0.08) & 18.32(0.01) & 16.75(0.01) &    0.11 &           C11 \\
 B174-G226 & -1.0 $\pm$ 0.2 &           C11 & 17.79(0.02) & 17.45(0.04) & 16.63(0.01) & 14.91(0.01) &    0.28 &           C11 \\
 B177-G228 & -1.2 $\pm$ 0.2 &           C11 & 19.79(0.04) & 19.52(0.08) & 18.93(0.01) & 17.43(0.01) &    0.18 &           C11 \\
 B178-G229 & -1.2 $\pm$ 0.2 &           C11 & 16.76(0.02) & 16.41(0.03) & 15.89(0.01) & 14.49(0.01) &    0.10 &           C11 \\
 B179-G230 & -1.0 $\pm$ 0.2 &           C11 & 17.22(0.02) & 16.85(0.04) & 16.18(0.01) & 14.72(0.01) &    0.10 &           C11 \\
 B180-G231 & -0.9 $\pm$ 0.2 &           C11 & 17.94(0.04) & 17.61(0.04) & 16.83(0.01) & 15.26(0.01) &    0.19 &           C11 \\
 B181-G232 & -0.5 $\pm$ 0.2 &           C11 & 18.95(0.03) & 18.64(0.06) & 17.65(0.01) & 15.98(0.01) &    0.13 &           C11 \\
 B182-G233 & -1.0 $\pm$ 0.2 &           C11 & 17.42(0.02) & 17.14(0.04) & 16.34(0.01) & 14.65(0.01) &    0.33 &           C11 \\
 B183-G234 & -0.5 $\pm$ 0.2 &           C11 & 18.13(0.02) & 17.84(0.04) & 16.85(0.01) & 15.18(0.01) &    0.15 &           SFD \\
 B184-G236 &  0.1 $\pm$ 0.2 &           C11 & 19.77(0.03) & 19.56(0.08) & 18.21(0.01) & 16.40(0.01) &    0.27 &           C11 \\
      B186 &  0.2 $\pm$ 0.3 &           C11 & 21.19(0.17) & 20.93(0.19) & 19.58(0.02) & 17.71(0.01) &    0.25 &           C11 \\
 B187-G237 & -1.6 $\pm$ 0.3 &           C11 & 19.23(0.03) & 18.95(0.07) & 18.15(0.02) & 16.36(0.01) &    0.36 &           C11 \\
      B274 & -0.9 $\pm$ 0.2 &           C\&R16 & 20.95(0.07) & 20.61(0.15) & 19.70(0.03) & 17.93(0.03) &    0.27 &           K12 \\
 B189-G240 &  0.4 $\pm$ 0.2 &           C11 & 19.73(0.06) & 19.70(0.15) & 18.15(0.02) & 16.35(0.01) &    0.24 &           C11 \\
 B190-G241 & -1.2 $\pm$ 0.2 &           C11 & 18.66(0.02) & 18.36(0.05) & 17.69(0.01) & 16.13(0.01) &    0.25 &           C11 \\
 B194-G243 & -1.4 $\pm$ 0.2 &           C11 & 18.66(0.02) & 18.32(0.05) & 17.83(0.02) & 16.43(0.02) &    0.19 &           C11 \\
 B193-G244 & -0.1 $\pm$ 0.2 &           C11 & 17.73(0.02) & 17.48(0.04) & 16.28(0.01) & 14.57(0.01) &    0.23 &           C11 \\
    SK072A &  0.0 $\pm$ 0.2 &           C\&R16 & 19.71(0.04) & 19.51(0.09) & 18.20(0.01) & 16.43(0.01) &    0.13 &           C11 \\
B103D-G245 & -0.8 $\pm$ 0.2 &           C11 & 19.55(0.04) & 19.15(0.07) & 18.34(0.01) & 16.74(0.01) &    0.18 &           C11 \\
 B472-D064 & -1.0 $\pm$ 0.2 &           C11 & 16.82(0.02) & 16.45(0.03) & 15.85(0.01) & 14.41(0.01) &    0.12 &           C11 \\
 B197-G247 & -0.3 $\pm$ 0.2 &           C11 & 20.03(0.05) & 19.79(0.10) & 18.56(0.01) & 16.74(0.01) &    0.30 &           C11 \\
 B198-G249 & -1.0 $\pm$ 0.2 &           C11 & 19.62(0.04) & 19.25(0.07) & 18.55(0.01) & 17.08(0.02) &    0.22 &           C11 \\
      B200 & -1.4 $\pm$ 0.3 &           C11 & 20.48(0.07) & 20.25(0.15) & 19.50(0.02) & 17.87(0.02) &    0.34 &           C11 \\
 B202-G251 & -1.2 $\pm$ 0.2 &           C11 & 19.47(0.03) & 19.17(0.06) & 18.50(0.01) & 16.99(0.01) &    0.10 &           SFD \\
 B204-G254 & -0.7 $\pm$ 0.2 &           C11 & 17.58(0.02) & 17.22(0.04) & 16.33(0.01) & 14.77(0.01) &    0.13 &           C11 \\
 B205-G256 & -0.9 $\pm$ 0.2 &           C11 & 17.28(0.02) & 16.92(0.04) & 16.26(0.01) & 14.80(0.01) &    0.12 &           C11 \\
 B206-G257 & -1.1 $\pm$ 0.2 &           C11 & 16.80(0.03) & 16.47(0.03) & 15.84(0.01) & 14.41(0.01) &    0.10 &           C11 \\
B110D-V296 & -1.4 $\pm$ 0.3 &           C11 & 20.06(0.05) & 19.76(0.10) & 19.09(0.01) & 17.54(0.01) &    0.28 &           C11 \\
 B207-G258 & -1.3 $\pm$ 0.2 &           C11 & 18.82(0.02) & 18.51(0.05) & 18.01(0.01) & 16.64(0.01) &    0.14 &           SFD \\
 B208-G259 & -0.4 $\pm$ 0.2 &           C11 & 20.05(0.05) & 19.73(0.10) & 18.65(0.01) & 16.93(0.01) &    0.21 &           C11 \\
      M009 & -1.8 $\pm$ 0.3 &           C11 & 19.24(0.03) & 18.95(0.07) & 18.51(0.01) & 17.12(0.01) &    0.12 &           C11 \\
 B209-G261 & -1.0 $\pm$ 0.2 &           C11 & 18.20(0.02) & 17.85(0.04) & 17.23(0.01) & 15.81(0.01) &    0.13 &           C11 \\
 B211-G262 & -1.4 $\pm$ 0.2 &           C11 & 18.24(0.02) & 17.88(0.04) & 17.41(0.01) & 16.04(0.01) &    0.14 &           C11 \\
 B212-G263 & -1.7 $\pm$ 0.2 &           C11 & 16.99(0.02) & 16.71(0.03) & 16.26(0.01) & 14.92(0.01) &    0.12 &           SFD \\
 B213-G264 & -0.8 $\pm$ 0.2 &           C11 & 18.80(0.02) & 18.44(0.05) & 17.55(0.01) & 15.92(0.01) &    0.17 &           C11 \\
 B214-G265 & -1.3 $\pm$ 0.2 &           C11 & 19.00(0.03) & 18.66(0.06) & 18.19(0.01) & 16.77(0.01) &    0.15 &           C11 \\
 B215-G266 & -0.6 $\pm$ 0.2 &           C11 & 19.32(0.03) & 18.98(0.06) & 17.98(0.01) & 16.35(0.01) &    0.14 &           C11 \\
 B217-G269 & -0.8 $\pm$ 0.2 &           C11 & 18.53(0.02) & 18.11(0.05) & 17.32(0.01) & 15.74(0.01) &    0.13 &           C11 \\
      M019 & -1.4 $\pm$ 0.2 &           C11 & 19.64(0.04) & 19.28(0.08) & 18.89(0.01) & 17.51(0.01) &    0.12 &           C11 \\
 B218-G272 & -0.8 $\pm$ 0.2 &           C11 & 16.67(0.02) & 16.33(0.03) & 15.59(0.01) & 14.08(0.01) &    0.17 &           C11 \\
 B220-G275 & -1.5 $\pm$ 0.3 &           C11 & 18.50(0.02) & 18.16(0.04) & 17.66(0.01) & 16.29(0.01) &    0.06 &           F12 \\
 B221-G276 & -1.1 $\pm$ 0.2 &           C11 & 18.61(0.02) & 18.29(0.05) & 17.58(0.01) & 16.01(0.01) &    0.26 &           C11 \\
 B224-G279 & -1.5 $\pm$ 0.2 &           C11 & 17.72(0.02) & 17.36(0.04) & 16.92(0.01) & 15.60(0.01) &    0.07 &           F12 \\
 B279-D068 &  0.0 $\pm$ 0.2 &           C11 & 20.77(0.07) & 20.78(0.19) & 19.28(0.02) & 17.43(0.01) &    0.29 &           C11 \\
 B225-G280 & -0.5 $\pm$ 0.2 &           C11 & 16.18(0.02) & 15.86(0.03) & 14.89(0.01) & 13.29(0.01) &    0.05 &           F12 \\
 B228-G281 & -0.7 $\pm$ 0.2 &           C11 & 18.93(0.03) & 18.56(0.05) & 17.61(0.01) & 15.92(0.01) &    0.24 &           C11 \\
 B229-G282 & -2.1 $\pm$ 0.3 &           C11 & 18.66(0.02) & 18.36(0.05) & 17.98(0.01) & 16.67(0.01) &    0.11 &           C11 \\
 B231-G285 & -1.0 $\pm$ 0.2 &           C11 & 18.96(0.03) & 18.58(0.05) & 17.91(0.01) & 16.45(0.01) &    0.15 &           C11 \\
 B233-G287 & -1.1 $\pm$ 0.2 &           C11 & 17.54(0.02) & 17.16(0.04) & 16.54(0.01) & 15.09(0.01) &    0.10 &           F12 \\
 B234-G290 & -0.8 $\pm$ 0.2 &           C11 & 18.72(0.02) & 18.32(0.05) & 17.48(0.01) & 15.94(0.01) &    0.23 &           C11 \\
 B283-G296 & -0.8 $\pm$ 0.2 &           C11 & 19.92(0.03) & 19.52(0.07) & 18.68(0.01) & 17.05(0.01) &    0.13 &           SFD \\
 B235-G297 & -0.9 $\pm$ 0.2 &           C11 & 18.15(0.02) & 17.80(0.04) & 17.10(0.01) & 15.53(0.01) &    0.14 &           C11 \\
 B237-G299 & -1.8 $\pm$ 0.2 &           C11 & 18.91(0.02) & 18.60(0.05) & 18.15(0.01) & 16.75(0.01) &    0.15 &           SFD \\
 B370-G300 & -1.6 $\pm$ 0.2 &           C11 & 17.87(0.02) & 17.62(0.04) & 17.00(0.01) & 15.45(0.01) &    0.27 &           C11 \\
 B238-G301 & -0.7 $\pm$ 0.2 &           C11 & 18.56(0.02) & 18.16(0.04) & 17.34(0.01) & 15.78(0.01) &    0.11 &           SFD \\
  B239-M74 & -0.9 $\pm$ 0.2 &           C11 & 18.84(0.02) & 18.50(0.05) & 17.81(0.01) & 16.32(0.01) &    0.12 &           C11 \\
 B240-G302 & -1.5 $\pm$ 0.2 &           C11 & 16.88(0.02) & 16.53(0.03) & 16.04(0.01) & 14.68(0.01) &    0.14 &           F12 \\
      B287 & -1.4 $\pm$ 0.4 &           C11 & 19.77(0.03) & 19.47(0.07) & 19.12(0.01) & 17.80(0.02) &    0.07 &           C11 \\
    SK104A & -0.8 $\pm$ 0.2 &           C\&R16 & 19.91(0.04) & 19.60(0.08) & 18.63(0.01) & 16.99(0.01) &    0.32 &           K12 \\
  V129-BA4 & -1.4 $\pm$ 0.2 &           C11 & 18.52(0.02) & 18.22(0.04) & 17.73(0.01) & 16.36(0.01) &    0.18 &           C11 \\
 B375-G307 & -0.9 $\pm$ 0.2 &           C11 & 19.46(0.03) & 19.07(0.06) & 18.31(0.01) & 16.77(0.01) &    0.21 &           C11 \\
     B270D & -2.1 $\pm$ 0.4 &           C\&R16 & 19.57(0.03) & 19.29(0.06) & 18.81(0.01) & 17.40(0.01) &    0.09 &           SFD \\
 B378-G311 & -1.6 $\pm$ 0.3 &           C11 & 19.00(0.02) & 18.71(0.06) & 18.25(0.01) & 16.93(0.01) &    0.13 &           C11 \\
 B381-G315 & -1.1 $\pm$ 0.2 &           C11 & 17.70(0.02) & 17.31(0.04) & 16.63(0.01) & 15.13(0.01) &    0.10 &           SFD \\
 B382-G317 & -1.9 $\pm$ 0.2 &           C11 & 18.77(0.02) & 18.43(0.04) & 17.97(0.01) & 16.63(0.01) &    0.11 &           SFD \\
 B383-G318 & -0.6 $\pm$ 0.2 &           C11 & 17.80(0.02) & 17.44(0.04) & 16.53(0.01) & 14.89(0.01) &    0.10 &           SFD \\
 B391-G328 & -1.2 $\pm$ 0.2 &           C11 & 19.01(0.02) & 18.64(0.05) & 17.99(0.01) & 16.53(0.01) &    0.08 &           SFD \\
 B393-G330 & -0.9 $\pm$ 0.3 &           C11 & 19.10(0.03) & 18.70(0.05) & 18.00(0.01) & 16.51(0.01) &    0.08 &           SFD \\
 B397-G336 & -1.2 $\pm$ 0.2 &           C11 & 18.06(0.02) & 17.74(0.04) & 17.15(0.01) & 15.74(0.01) &    0.08 &           SFD \\
 B398-G341 & -0.4 $\pm$ 0.3 &           C11 & 19.62(0.03) & 19.35(0.06) & 18.35(0.01) & 16.68(0.01) &    0.07 &           SFD \\
 B399-G342 & -1.7 $\pm$ 0.6 &           C11 & 18.90(0.02) & 18.59(0.04) & 18.17(0.01) & 16.88(0.01) &    0.08 &           SFD \\
 B401-G344 & -1.9 $\pm$ 0.4 &           C11 & 18.25(0.02) & 17.98(0.04) & 17.62(0.01) & 16.38(0.01) &    0.08 &           SFD \\
\end{longtable}
\label{tab:phot}
\tablefoot{A static aperture of 3 arcseconds was used for all GCs and the magnitudes were \textbf{not} aperture corrected, under the assumption that GCs show no colour gradients. Therefore the noted magnitudes do not correspond to the total magnitude for the GCs and can only be used for colours. C11: \cite{Caldwell2011}; C\&R16: \cite{Caldwell2016}; L20: \cite{Larsen2020};  F10: \cite{Fan2010}; F12: \cite{Federici2012}; K12: \cite{Kang2012}; L21: \cite{Larsen2021}; SFD: \cite{Schlegel1998} dust map with the correction of \cite{Schlafly2010}}
\end{appendix}
\label{LastPage}
\end{document}